\documentclass[twocolumn,trackchanges]{aastex62}
\usepackage[T1]{fontenc}
\usepackage[latin1]{inputenc}
\usepackage{enumerate,times,subfigure,multirow,xcolor} 
\usepackage{hyperref,astrojournals,amsmath,amssymb,mathtools,graphicx,txfonts,microtype,nicefrac,xspace}


\newcommand{\Md}{M_{\rm disk}}
\newcommand{\Mbh}{M_{\rm \bullet}}

\newcommand{\arcs}{\mbox{\ensuremath{^{\prime\prime}}}}

\renewcommand*{\vec}[1]{\boldsymbol{#1}}
\newcommand{\dg}{^\circ}

\shorttitle{Observing Eccentric Nuclear Disks}
\shortauthors{Wernke \& Madigan}

\begin{document}
\bibliographystyle{apj}

\title{Photometry and Kinematics of Self-Gravitating Eccentric Nuclear Disks}

\author{Heather N. Wernke}
\affiliation{JILA and Department of Astrophysical and Planetary Sciences, CU Boulder, Boulder, CO 80309, USA}
\email{heather.wernke@colorado.edu}

\author{Ann-Marie Madigan}
\affiliation{JILA and Department of Astrophysical and Planetary Sciences, CU Boulder, Boulder, CO 80309, USA}

\begin{abstract}
The Andromeda Galaxy hosts an elongated nucleus with (at least) two distinct brightness peaks.  The double nucleus can be explained by the projection of a thick, apsidally-aligned eccentric nuclear disk of stars in orbit about the central black hole.  Several nearby early-type galaxies have similar asymmetric nuclear features, indicating the possible presence of eccentric nuclear disks.  
We create simulated photometric (surface density) and kinematic (line-of-sight velocity) maps of eccentric nuclear disks using $N$-body simulations.  We image our simulations from various lines of sight in order to classify them as double nuclei, offset nuclei, and centered nuclei.  We explore the effects of mass segregation on the photometric maps, finding that heavier stars are concentrated in the brighter peak. The average line-of-sight velocity values are lower in an eccentric nuclear disk than for a circular ring about the supermassive black hole. The velocity dispersion values are higher and peak at the position of the supermassive black hole, which does not typically match the peak in photometry. 
\end{abstract}


\section{INTRODUCTION}  
\label{sec:intro}  

The elliptical nucleus of M31 was first resolved by the balloon-borne Stratoscope II telescope, which showed an asymmetric nucleus.  Later, \citep{Nieto1986} showed that the nucleus was not only asymmetric, but offset from the bulge.  Almost 20 years after the initial observations,
Hubble Space Telescope Wide Field/Planetary Camera ({\it HST}/WFPC) images, with $0\farcs043$ per pixel resolution, showed that the nucleus of M31 contains two separate components.  The component with the higher surface brightness, which was observed by Stratoscope II, is known as P1; the fainter peak is designated P2.  P1 and P2 are separated by $0\farcs49\pm0\farcs01$ or 1.8 pc \citep{Lauer1993}. P2 lies closest to the nuclear/bulge center which is coincident with the peak in stellar velocity dispersion. 
That is to say that the kinematic center is separate from the luminosity peak \citep{Dressler1988, Kormendy1988}. 
P1 and P2 contain stars of the same stellar spectral type (K), indicating that they are a part of the same system \citep{Kormendy1999}.      
\citet{Tremaine1995} explained the double nucleus of M31 as the projected appearance of an apsidally-aligned eccentric stellar disk.  The stars in the eccentric nuclear disk travel around the supermassive black hole within the radius of influence on Keplerian orbits.  The two observed components correspond to apocenter (P1) and pericenter (P2) of the eccentric nuclear disk.

The fact that we see an eccentric nuclear disk in our largest galactic neighbor suggests that apsidally-aligned stellar disks may be a common occurrence in galactic nuclei. 
Even with observational challenges due to resolution, evidence may exist for many eccentric nuclear disks in the local universe.  \citet{Lauer1996} showed that NGC 4464B also contains a double nucleus and is likely host to a similar structure to the eccentric disk in M31.  \citet{Lauer2002} identified numerous galaxies with local surface brightness minima in their centers, noting that at least some may be related to the double-nucleus systems.  \citet{Lauer2005} presented {\it HST} observations of 77 early-type galaxies. They noted that the galaxies with offset centers may be poorly resolved examples of double nuclei or central minima.  About 15\% of nearby early-type galaxies have features consistent with eccentric nuclear disks seen from different orientations on the sky \citep{Lauer2005}. 
Furthermore, \citet{Gruzinov2020} showed that a lopsided configuration can be an equilibrium mode of a rotating nuclear star cluster. 

An eccentric nuclear disk is composed of stars moving on apsidally aligned, near-Keplerian orbits around a central supermassive black hole.  \citet{Hop10a,Hop10b} showed that massive eccentric nuclear disks can form from gas-rich galaxy mergers.  In a galaxy with a scoured stellar core due to the binary black hole inspiral \citep{Beg80}, apsidal precession will be dominated by the gravitational potential of the newly formed disk and not by a spherical background potential. This increases the chances of the eccentric nuclear disk being stable. 
The presence of an eccentric nuclear disk in a galaxy could therefore provide information about the galaxy's merger history. 
Stellar orbits in the disk precess prograde with respect to their orbital angular momenta. Higher eccentricity orbits precess more slowly and end up behind the bulk of the disk, which then torques them to lower eccentricities. The orbits then precess more rapidly. In this way, orbits precess back and forth across the disk oscillating in eccentricity.   
This secular mechanism may explain the high rate of tidal disruption events (TDEs) observed in post-merger, starburst K+A/E+A galaxies \citep{Madigan2018}.  Stars from these disks preferentially disrupt at orbital inclinations of $0^\circ$ and $180^\circ$ \citep{Wernke2019}. Combined with the high rate of TDE production, this creates the conditions for overlapping TDE disks.  The high TDE rates also create more opportunity for gravitational wave bursts from TDEs to be observed by the Laser Interferometer Space Antenna (LISA) \citep{Pfister2021}.  Eccentric nuclear disks should also efficiently torque orbits of compact stellar remnants to high eccentricities which may result in extreme mass ratio inspirals (EMRIs), another exciting target for LISA. \\

In this paper, we show that double nuclei, offset nuclei, and nuclei with central minima can all be recovered by viewing eccentric nuclear disks in isolation from different lines of sight. 
We also show that gravitational mass segregation leads to heavier stars concentrating in the apocenter peak. 
Photometric and spectroscopic maps of eccentric nuclear disks have been produced to compare the M31 eccentric disk by \citet{Lauer1993}, \citet{Tremaine1995}, \citet{Kormendy1999}, \citet{Pei03}, \citet{Bro13}, and \citet{Lockhart2018}. The novelty of the work presented here is that we "observe" eccentric nuclear disks in \textit{self-gravitating} $N$-body simulations. 
We quantify the expected prevalence of double nuclei and offset nuclei and compare to the \citet{Lauer2005} survey. 
We note that using $N$-body simulations restricts us to simulating a narrow range in semi-major axis. In effect we simulate only the innermost edge of the disk. For this reason we refrain from making direct comparisons with the M31 nucleus; rather we examine the qualitative features of the self-gravitating structure.  Furthermore, we simulate a disk that is a factor of one hundred lower in mass than the central black hole which means that our disks are photometrically modest systems. In reality, such a disk would be overwhelmed by background stellar cusps and bulges. However, we simulate the eccentric nuclear disks in isolation, without the presence of a bulge or nuclear cusp, in order to pinpoint the features made by this lopsided structure.

We present the paper in the following manner.  In Section~\ref{sec:method} we describe the initial conditions and parameters for our simulations, and explain the methods used to rotate and "observe" our simulated disks.  In Section~\ref{sec:map} we explore the photometric maps of eccentric nuclear disks from different orientations.  We also explore the effects of differing resolution and number of stellar particles.  In Section~\ref{sec:vel}, we make  kinematic maps of an eccentric nuclear disk, including line-of-sight velocity and velocity dispersion maps.  In Section~\ref{sec:disc} we summarize and discuss our results.

\section{METHODS}
\label{sec:method}

We run $N$-body simulations of eccentric nuclear disks about supermassive black holes with {\tt REBOUND} \citep{Rein2012} and the {\tt IAS15} integrator \citep{Rein2015}.  We initialize simulations with the following parameters: $N=100-1000$ star particles, each with an initial eccentricity of $e = 0.8$, a range of semi-major axes ($a=1-2$) with a surface density of $\Sigma\propto a^{-2}$, Rayleigh distributed inclinations with mean $10\dg$, Gaussian distributed longitude of pericenter ($\varpi$) with mean 1 radian and standard deviation 0.5 radian, and a disk mass of $10^{-2}\Mbh$, where $\Mbh$ is the mass of the black hole. The particles are distributed randomly between $[0, 360\dg)$ in mean anomaly ($M$) and longitude of ascending node ($\Omega$). 

At the end of each integration, we populate each orbit with 100 stars uniformly in mean anomaly; this increases the effective resolution.  We then make surface density plots for the eccentric nuclear disk looking down the positive $z$-axis, rotating the disk itself to produce different lines of sight.  These orientations are generated by points randomly drawn on a sphere, represented by
\begin{subequations}
    \begin{align}
        \theta &= \arccos(2u_1 -1) \\
        \phi &= 2\pi u_2,
    \end{align}
    \label{eq:sphere}
\end{subequations}
where $u_1$ and $u_2$ are both uniform in $(0,1)$.  We use $\hat{k}$ to describe the unit vector pointing to $(\theta,\phi)$.  We rotate the angular momentum vector of the disk to the chosen point on the sphere by rotating by an angle $\omega$ around the vector $v_{\rm rot}$ with Rodrigues' rotation formula, where
\begin{subequations}
    \begin{align}
        \omega &= \arccos(\hat{z} \cdot \hat{k})
        \\
        v_{\rm rot} &= \hat{z} \times \hat{k}.
        \label{eq:rot}
    \end{align}
\end{subequations}
With this definition, $\theta=90\dg$ indicates that the disk will be seen edge on.  $\phi$ then determines if the disk is viewed more along the major axis or the minor axis.  $\phi=0\dg$ or $180\dg$ means we look down the major axis, while $\phi=90\dg$ or $270\dg$ means we look down the minor axis of the disk.  $\theta=0\dg$ or $180\dg$ means we will observe a face on disk. 

With 100 different "observations" of each disk, following \citet{Lauer2005} we classify them by eye into three main types of nuclei: double nuclei or nuclei with central minima, offset nuclei, and normal or typical nuclei.  We refer to these categories as double nuclei, offset nuclei, and centered nuclei respectively.

In addition to photometric maps, we create line-of-sight velocity maps, velocity dispersion maps, and maps of the skew and kurtosis in the line-of-sight velocity in each orientation.  
We simulate a circular disk with the same parameters as our eccentric nuclear disk simulations to demonstrate the differences between them.  We note that these maps and categories are illustrative only as the photometric and kinematic appearance of an eccentric nuclear disk (including the basic question of which of P1 and P2 is brighter) will be wavelength dependent in reality. Disks of various ages, mass functions, and degrees of mass segregation may present themselves very differently as a function of photometric band.

\begin{figure}[t!]
    \centering
    \includegraphics[trim=0.75cm 0.75cm 2cm 1.75cm, clip=true, width=\columnwidth]{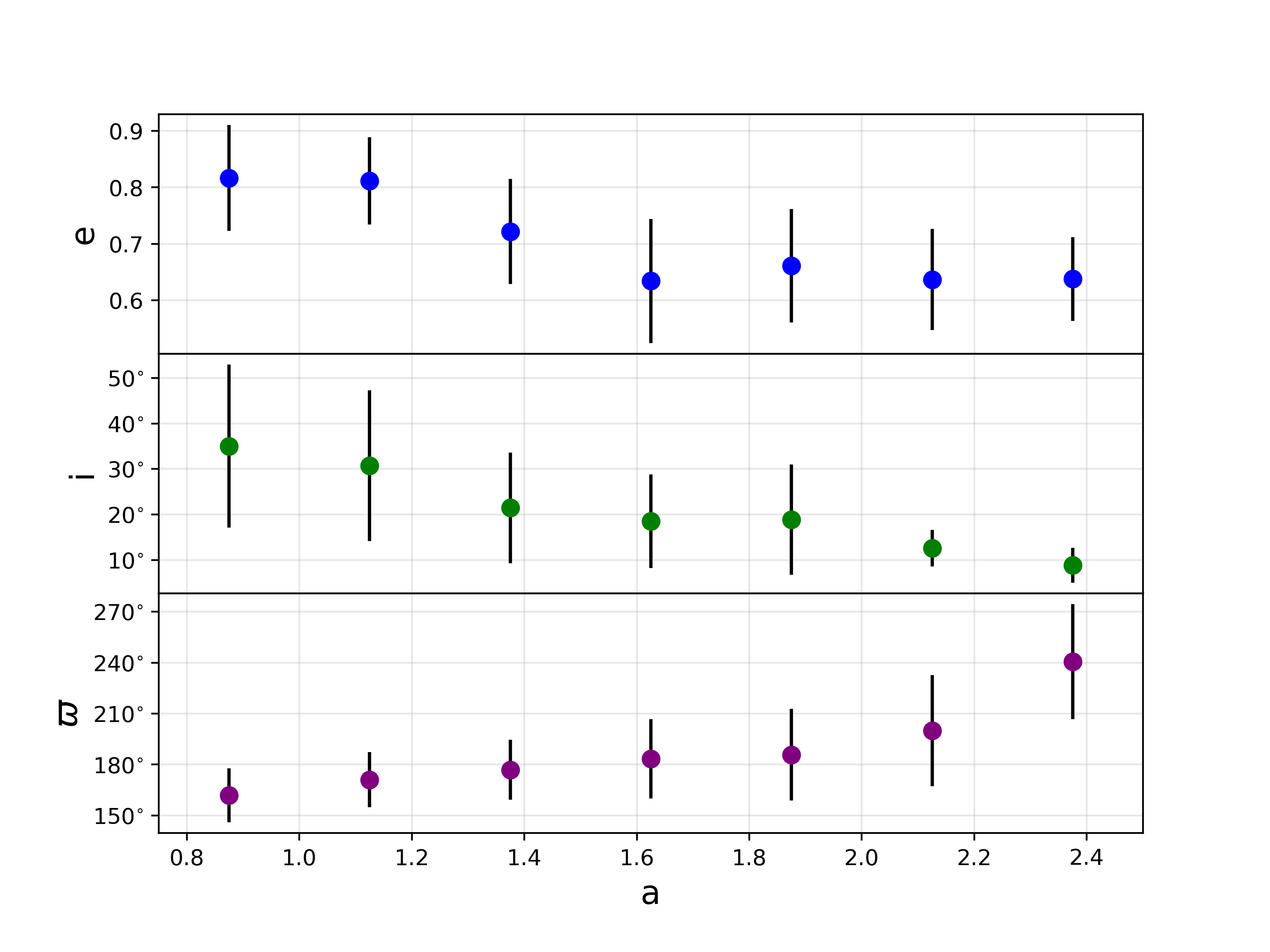}
    \caption{Orbital elements of an eccentric nuclear disk at 200 orbital periods (two secular times) as a function of semi-major axis: eccentricity (top), inclination (middle), and longitude of pericenter (bottom); mean values with $1\sigma$ error bars.}
    \label{fig:orbits}
\end{figure}

\section{PHOTOMETRIC MAPS}
\label{sec:map}

We create photometric maps of our simulated eccentric nuclear disks after two dynamical times.  Here we define the secular dynamical timescale as
\begin{equation}
t_{\rm sec}\equiv\left(\frac{\Mbh}{\Md}\right)P,
\end{equation}
where $\Md$ is the mass of the disk and $P$ is the orbital period for a star at the inner edge of the disk \citep{Rauch1996}.  In our simulations, $t_{\rm sec}=100P$.  We create photometric maps after two dynamical times to allow the disk more than a full secular time to relax. In particular the disk moves away from its (arbitrary) initial conditions and develops a negative eccentricity gradient, in which stars at lower semi-major axes have higher equilibrium eccentricities. Furthermore, \citet{Foote2020} find that vertical mass segregation occurs within two dynamical times (see section \ref{ss:mass-seg}).
Figure~\ref{fig:orbits} shows the orbital eccentricity (top), inclination (middle), and longitude of pericenter (bottom) of a representative eccentric nuclear disk at two secular times as a function of semi-major axis.

\begin{figure*}
    \centering
    \includegraphics[trim=3.25cm 3.75cm 6cm 4.25cm, clip=true, width=\textwidth]{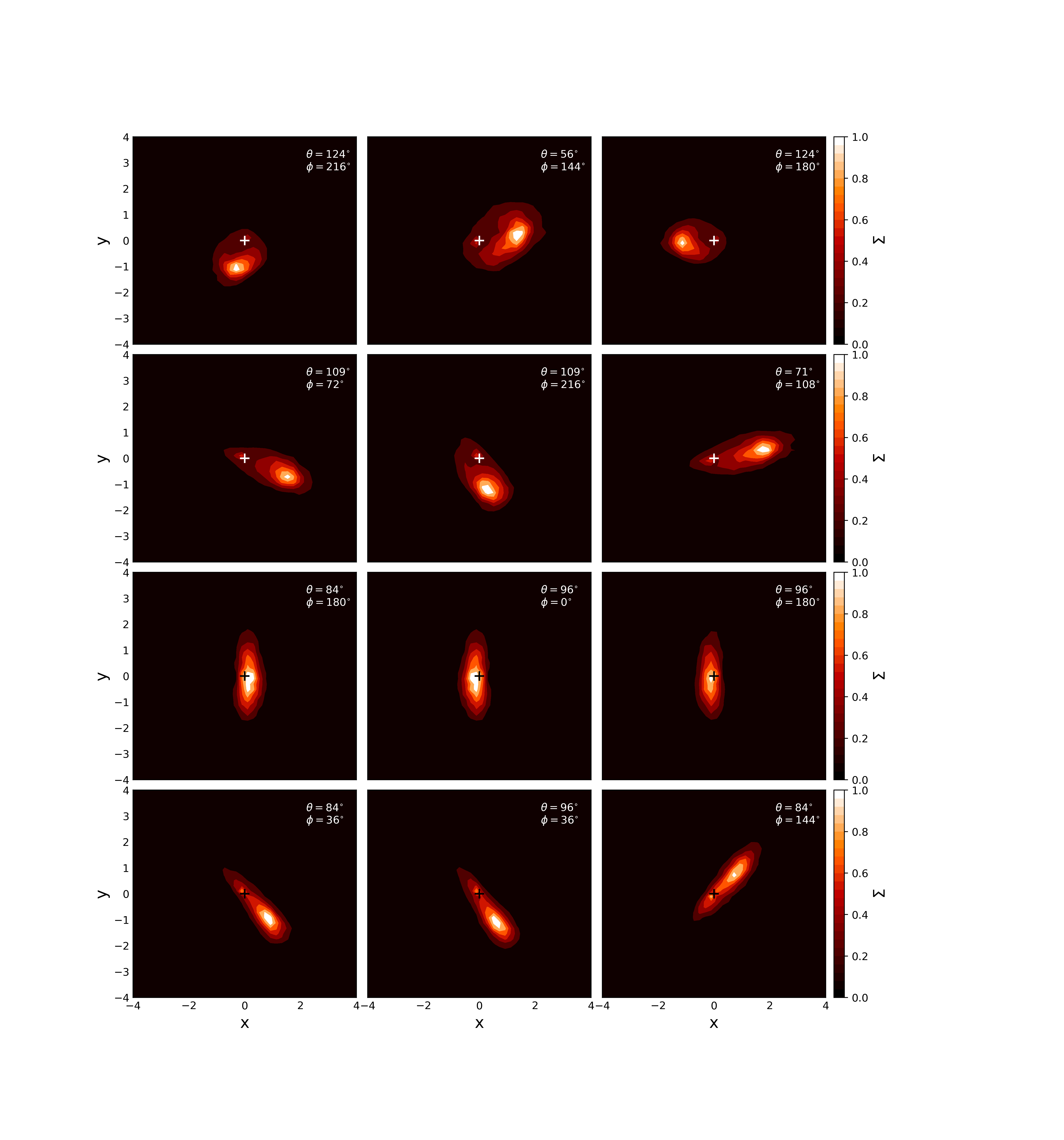}
    \caption{A sample of photometric maps of an eccentric nuclear disk viewed from different orientations.  The position of the supermassive black hole is marked with a cross at the origin.  This sample was created from an $N$-body simulation with $N=1000$ stars and a photometric resolution of $0.2$.  The first two rows show examples of offset nuclei.  The third row shows examples of centered nuclei, while the bottom row shows examples of double nuclei.  $\theta$ and $\phi$, randomly selected points on an sphere that generate new disk orientations as given in Equation \ref{eq:sphere}, are listed for each panel. 
    }
    \label{fig:photo-map}
\end{figure*}

In Figure~\ref{fig:photo-map} we show a small sample of photometric maps of our simulated eccentric nuclear disk viewed from various angles.  The location of the supermassive black hole is indicated with a cross at the origin. Most lines of sight produce offset, elliptical nuclei, like those seen in the top two rows. Several of the surface density profiles reveal an M31-like tail or faint second peak.  Some of these M31-like double nuclei are shown in the last row of Figure~\ref{fig:photo-map}. Finally, we even see some nuclei that appear photometrically "normal," centered on the supermassive black hole, like those seen in the third row of Figure~\ref{fig:photo-map}. The surface density contours are still elliptical. 

With $N=1000$ star particles and a resolution of $0.2$ (where $a = 1$ is the semi-major axis at the inner edge of the disk),  78\% of the orientations result in an elliptical nucleus that is offset from the central black hole.  
A strong double nucleus occurs 16\% of the time and 6\% of the orientations result in a single peak that is centered on the black hole. 
The percentages of double, centered, and offset nuclei for varying resolutions and $N$ are listed in Table~\ref{tab:percent}.  
\begin{table}[t]
\centering
\caption{Percentage of double nuclei, centered nuclei, and offset nuclei seen with varying resolution and $N$ star particles.  Resolution is presented in terms of the inner edge of the disk.  The sample size for each is 100 orientations.}
\begin{tabular}[t]{c c|c c c}
 Resolution & N & Double & Centered & Offset\\
 \hline
 $0.4$ & 100 & 0\% & 9\% & 91\%\\
 $0.2$ & 100 & 8\% & 7\% & 85\%\\
 $0.13$ & 100 & 15\% & 6\% & 79\%\\
 $0.4$ & 1000 & 1\% & 6\% & 93\%\\
 $0.2$ & 1000 & 16\% & 6\% & 78\%\\
 $0.13$ & 1000 & 34\% & 5\% & 61\%
\end{tabular}
\label{tab:percent}
\end{table}
\begin{figure*}
    \centering
    \includegraphics[trim=2.5cm 0cm 2.25cm 0.75cm, clip=true, width=\textwidth]{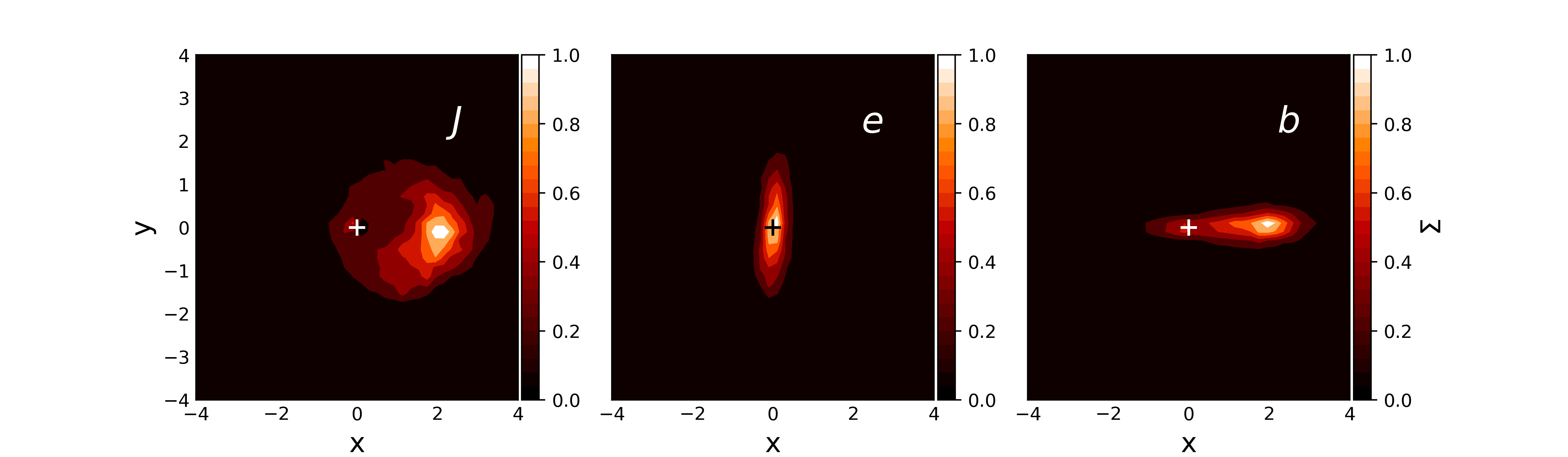}
    \caption{Photometric maps of an eccentric nuclear disk (N=1000 stars, $0.2$ resolution) viewed along the angular momentum vector ($\vec{J}$), eccentricity vector ($\vec{e}$), and semi-latus rectum ($\vec{b}$). The location of the supermassive black hole is marked with a cross at the origin.}
    \label{fig:specific-map}
\end{figure*}
We vary resolutions between $0.4$, $0.2$, and $0.13$ with both $N=100$ and $N=1000$ stars.  As expected, more double nuclei are seen with increasing resolution and numbers of stars.  For reference, the double nucleus of M31 was observed with a resolution of $0\farcs044$ per pixel \citep{Lauer1993}, or $0.75$ in code units if the inner edge of the disk is about 0.2pc ($a = 1$).  This adopts the distance to M31 of 770kpc \citep{Freedman1990} so that $1\arcs = 3.4$pc. 

\begin{figure}[ht!]
    \centering
    \includegraphics[trim=0cm 0cm 0.25cm 0.75cm, clip=true, width=\columnwidth]{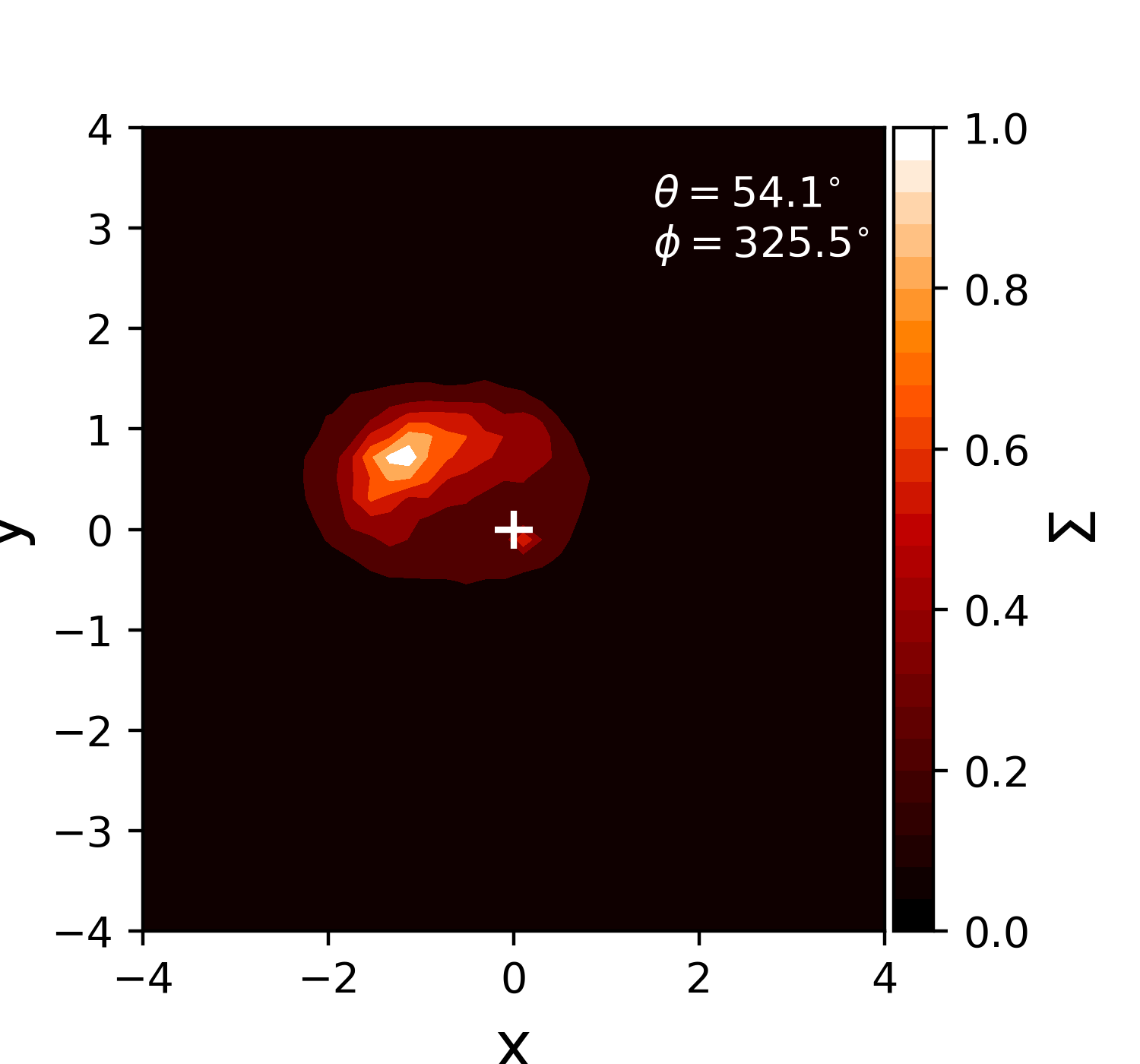}
    \caption{Photometric map of an eccentric nuclear disk with N=1000 stars and $0.2$ resolution matching the observed orientation of the disk in the M31 nucleus.  We use the angles given by \citet{Pei03} to define $\theta$ and $\phi$ for M31.  We do not observe the M31 disk edge-on, but it is sufficiently inclined to produce a small, second peak.}
    \label{fig:m31}
\end{figure}

The three main categories of nuclei are best captured by viewing specific orientations as shown in Figure~\ref{fig:specific-map}.  If we orient the disk such that its angular momentum vector points towards us, and the disk is observed face on, we see a maximally offset, elliptical nucleus.  The surface density contours become more elliptical and centered as the disk is rotated such that the eccentricity vector, or the major axis, points towards the observer. Looking down the eccentricity vector of the eccentric nuclear disk results in a single peak centered on the black hole.  These centered orientations may appear photometrically ``normal'' (albeit with highly elliptical contours), but we should still expect to see unusual velocity signatures (see Section \ref{sec:vel}).  Finally, if the observer looks along a line of sight parallel to the minor axis of the disk, they will observe a double-peaked nucleus similar to that found in M31.  Figure \ref{fig:m31} shows a photometric map that matches the observed orientation of the M31 double nucleus.  We use angles derived from the nonaligned model in \citet{Pei03} to define the $\theta$ and $\phi$ orientation angles for the M31 nuclear disk. We again note that there is no bulge or background stellar cusp in our simulations. The inclusion of such populations would enhance the observed luminosity of the secondary (P2) peak \citep{Tremaine1995}.

In general, we observe a centered nucleus when viewing the eccentric nuclear disk along its eccentricity vector, or major axis, $\pm6\dg$ in $\theta$ and $\pm12\dg$ in $\phi$.  Similarly, we observe a double nucleus when viewing the eccentric nuclear disk along its minor axis $\pm6\dg$ in $\theta$ and $\pm30\dg$ in $\phi$. In fact we find that the appearance of a double nucleus is most obvious when the disk is viewed just off of the minor axis, so that the observer's line of sight penetrates through more of the disk at pericenter, enhancing the fainter (P2) peak. 

\subsection{Mass Segregation}
\label{ss:mass-seg}

\begin{figure*}
    \centering
    \includegraphics[trim=2cm 1cm 2.5cm 2cm, clip=true, width=\textwidth]{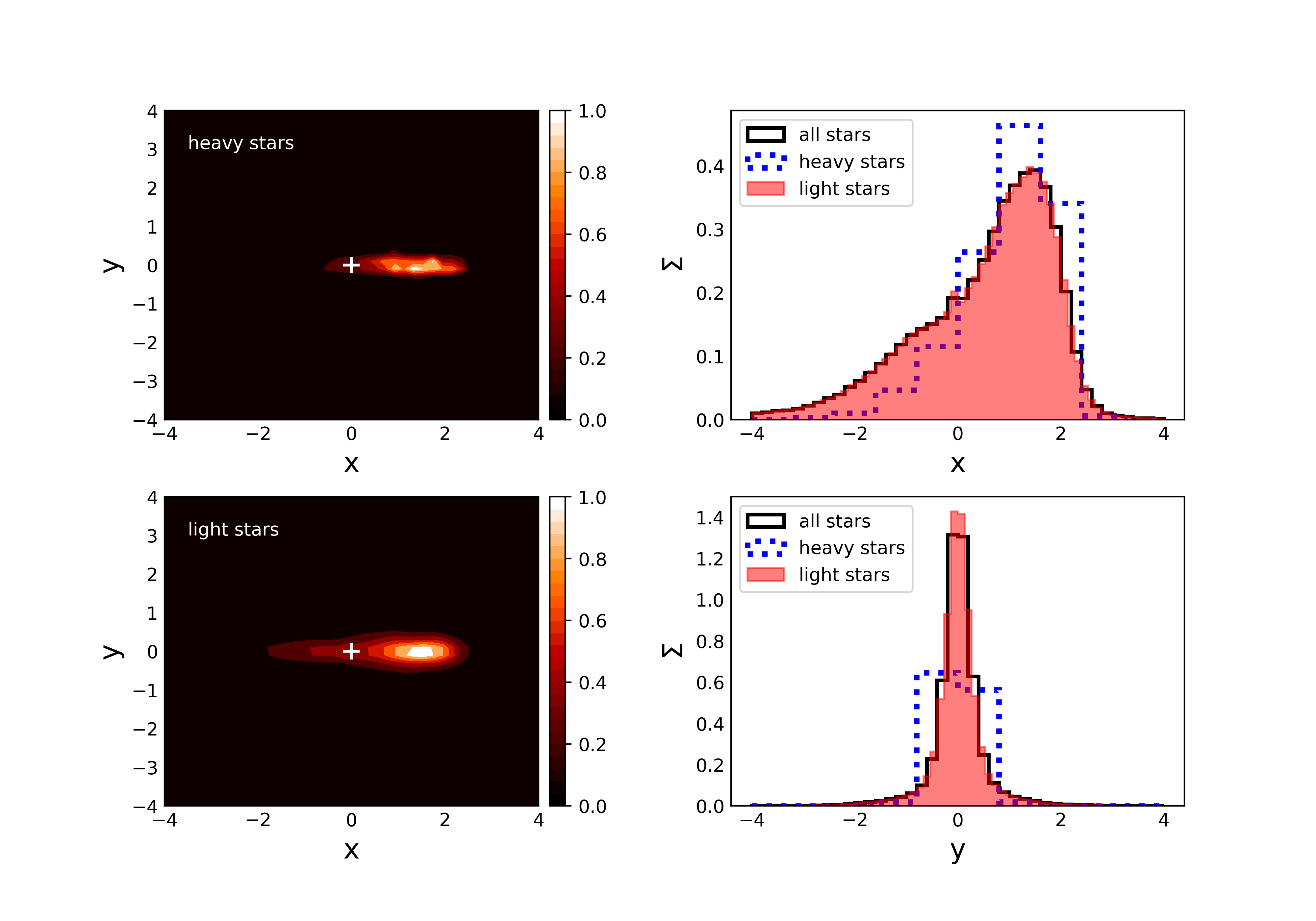}
    \caption{Mass segregation in photometric maps of simulated eccentric nuclear disks with $N$=400 light stars, $N$=5 heavy stars, and a resolution of $0.2$.  15 simulations from \citet{Foote2020} are combined to create this plot.  Top left: surface density plot of heavy stars.  The cross marks the position of the supermassive black hole.  Bottom left: surface density plot of light stars for the disk orientation.  The cross marks the position of the supermassive black hole.  Top right: x-histogram of the surface density plots (i.e., moving along the disk major axis).  The solid, black line histogram includes all stars.  The dotted, blue histogram shows heavy stars (at low semi-major axis).  The filled-in, red histogram shows light stars (at high semi-major axis).  Bottom right: the y-histogram of the surface density plots.  The density ratio between P1 and P2 is larger for the heavy star population than for the light star population. }
    \label{fig:mass-seg}
\end{figure*}

\citet{Foote2020} showed that high mass stars segregate to lower semi-major axes and inclinations in an eccentric nuclear disk.  In their $N$-body simulations, heavy stars differ in mass by a factor ten from light stars. This simple two-species model, put forward by \citep{Ale09}, approximates an evolved stellar population (coeval or continuously star-forming); light stars represent old low-mass main-sequence dwarfs, white dwarfs, and neutron stars with masses of order a solar mass, and heavy stars represent stellar-mass black holes with masses of order ten solar masses. 
Here we use simulations from \citet{Foote2020} in order to show the effect of mass segregation on the surface density of eccentric nuclear disks.

Here we show the results of simulations in the strong mass segregation regime \citep{Ale09}, where the heavy stars sink efficiently to lower semi-major axes and orbital inclination due to dynamical friction.  In Figure \ref{fig:mass-seg}, we combine 15 simulations, each with 400 light stars and 5 heavy stars. The relaxation coupling parameter for this model is $\Delta = 0.384$ \citep[][equation 10]{Ale09}. The heavy stars concentrate preferentially in the brighter peak (P1). We see the same effect in simulations with larger $\Delta$ values (weaker mass segregation) but to a lesser extent.  
While the observational effects of this mass segregation will depend on the wavelength of the observations and the age of the stellar population, our results indicate that the two peaks (P1 and P2) of an eccentric nuclear disk with an evolved stellar population will have differing mass to light ratios at all wavelengths as the fraction of heavy objects are enhanced in P1 relative to P2.  It makes sense that the heavy stars are found in the most offset peak because stars at low semi-major axes are the most eccentric (see Figure~\ref{fig:orbits} and \citet{Madigan2018}) and they spend the most time at apocenter.

\section{Kinematic Maps}
\label{sec:vel}

\begin{figure*}
    \centering
    \includegraphics[trim=2.75cm 2cm 2.5cm 3cm, clip=true, width=\textwidth]{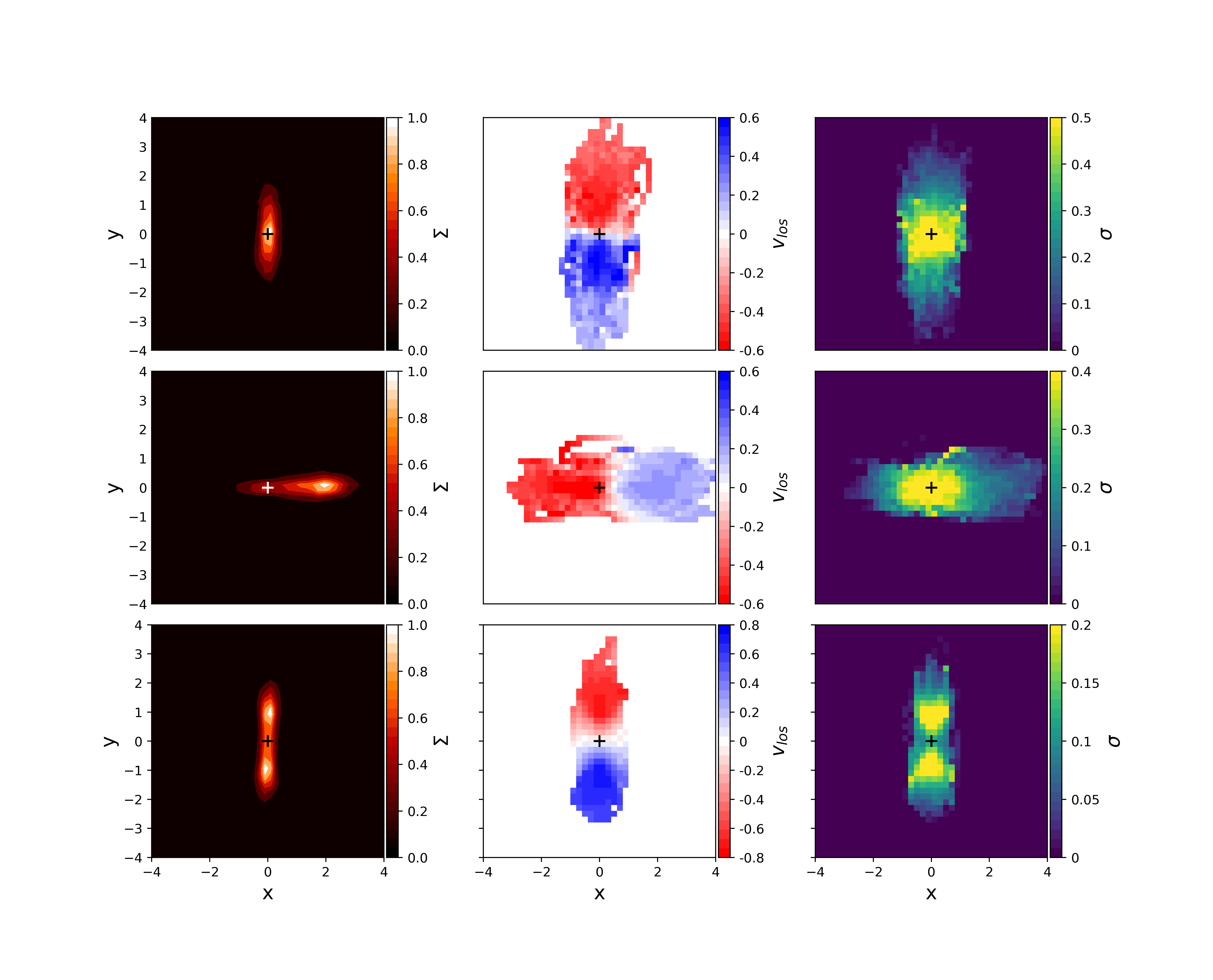}
    \caption{Top row: An eccentric nuclear disk with N=1000 stars and $0.2$ resolution viewed along its major axis.  Middle row:  An eccentric nuclear disk with N=1000 stars and $0.2$ resolution viewed along its minor axis.  Bottom row: A circular disk with N=1000 stars and resolution of $0.2$ viewed edge on.  The left column shows the surface density. The position of the supermassive black hole is marked with a cross.  The middle column shows the line-of-sight velocity. Positive velocity is out of the page, so that blue indicates blue-shifted stars and red indicates red-shifted stars.  The right column shows the standard deviation of the line-of-sight velocity.
    }
    \label{fig:los-vel}
\end{figure*}

In Figure \ref{fig:los-vel}, we show photometric maps of disks alongside their line-of-sight velocity and velocity dispersion maps.  The top row shows an eccentric nuclear disk oriented such that its eccentricity vector, or major axis, is pointed towards the observer.  The middle row has the disk's minor axis pointed towards the observer. The bottom row shows an edge-on circular ring for comparison. The ring is initialized with the same orbital parameters as the eccentric nuclear disk but with zero eccentricity.

In the top row, the photometric map shows what looks like an axisymmetric disk of stars orbiting a black hole (indicated by the cross at the origin).  The line-of-sight velocities, however, are lower than in the circular disk. 
This is due to the fact that stars spend most of the time at apocenter in an eccentric nuclear disk, where their velocity is lower than the circular velocity at that same radius.  The velocity dispersion values peak around the black hole in the eccentric nuclear disk, but are significantly larger than in the circular case.  

The peak of the photometric values in the middle row is clearly offset from the black hole, with a faint second peak. The blue-shifted (right) side of the disk corresponds to apocenter of the disk and has a much lower velocity than the red-shifted (left) side which maps to pericenter of the disk.  The black hole lies within the red-shifted side. 
The velocity dispersion peaks around the supermassive black hole, even though the luminosity in the surface density plot peaks to the right of the supermassive black hole.
In the circular disk, the strengths of the red- and blue-shifted velocities are equal in value and the black hole is directly in the center.  
 
In the bottom row, we show an edge-on circular ring instead of an eccentric nuclear disk.  When viewed edge-on, the circular ring of stars appears elongated and centered on the supermassive black hole, similar to the top row of Figure~\ref{fig:los-vel}.  The line-of-sight velocity distribution for the circular disk is symmetric about its rotation axis. The line-of-sight velocity dispersion is centered on the black hole with a central minimum, as expected for a circular ring of stars.

\begin{figure*}
    \centering
    \includegraphics[trim=2.75cm 2cm 2.5cm 3cm, clip=true, width=\textwidth]{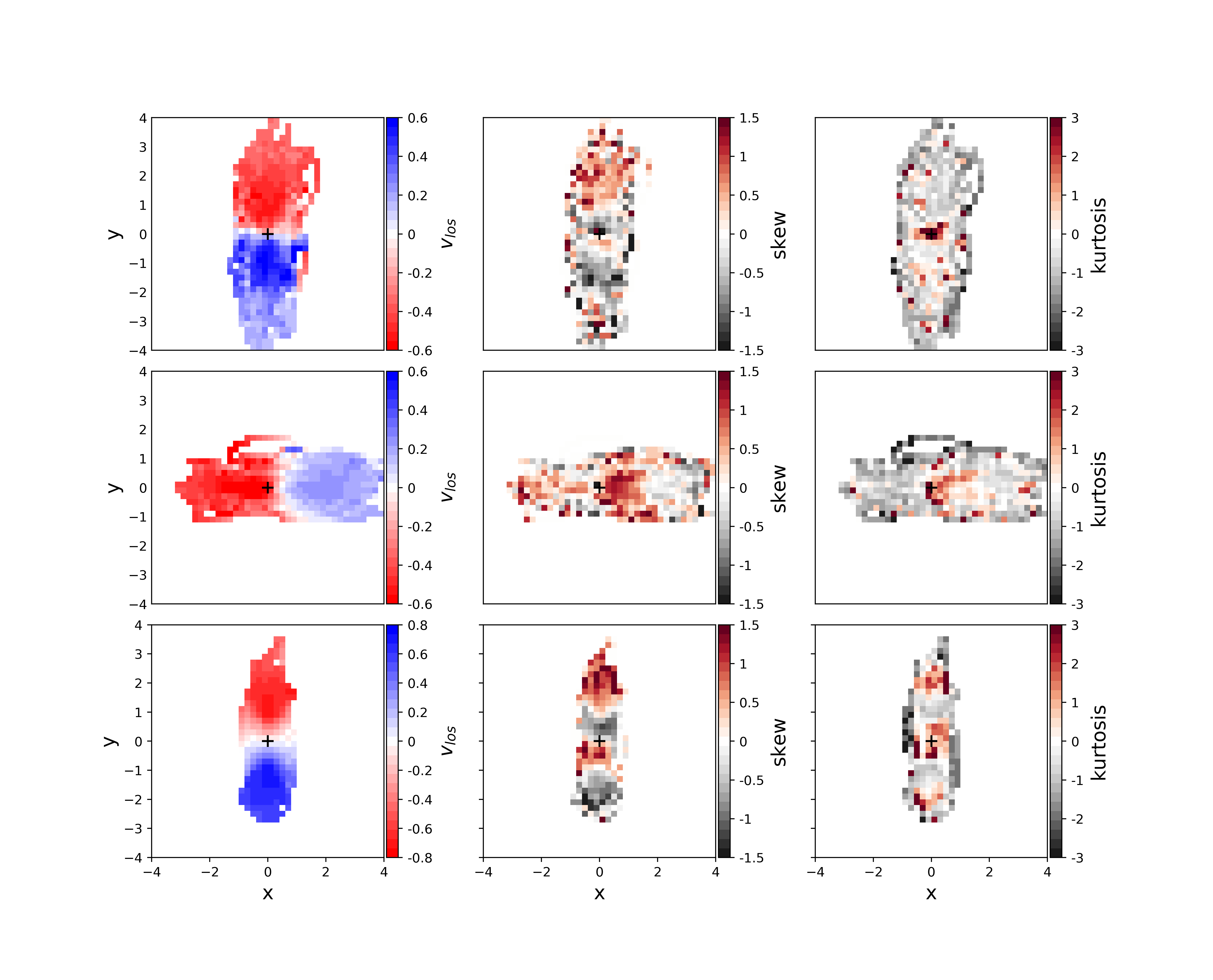}
    \caption{Top row: An eccentric nuclear disk with N=1000 stars and $0.2$ resolution viewed along its major axis.  Middle row:  The same eccentric nuclear disk viewed along its minor axis.  Bottom row: A circular ring viewed edge-on. The left column shows the line-of-sight velocity.  Positive velocity is out of the page, so that blue indicates blue-shifted stars and red indicates red-shifted stars.  The position of the supermassive black hole is marked with a cross.  The middle column shows the skew in the line-of-sight velocity and the right column shows the kurtosis in the line-of-sight velocity.
    }
    \label{fig:skew-kurt}
\end{figure*}

In Figure~\ref{fig:skew-kurt}, we take a closer look at the kinematic profiles of our eccentric nuclear disks by mapping the skew and kurtosis in the line-of-sight velocity distributions.  As in Figure~\ref{fig:los-vel}, the top row shows an eccentric nuclear disk oriented such that its eccentricity vector is pointed towards the observer.  The middle row shows an eccentric nuclear disk with its minor axis pointed towards the observer.  The bottom row shows an edge-on circular ring for comparison.

In the edge-on circular ring (bottom row), the outer regions of the disk ($a>1$), are skewed such that they have tails approaching zero velocity; this gives them positive and negative skew values for stars with negative and positive line-of-sight velocities respectively. Stars with the fastest line-of-sight velocities are in this region, and we look through a slower-moving portion of the disk.  Within the inner edge of the disk ($a<1$), the line-of-sight velocities are small, as we are observing stars with primarily transverse velocities.  The velocity distributions are quite flat in the central region with more low values than high to give skew values opposite to those in the outer disk.

Although noisier, the eccentric nuclear disk in the top row of Figure~\ref{fig:skew-kurt} shows very similar behavior in skew to the circular disk.  The inner regions with opposite skew values appear smaller due to the orientation of the eccentric orbits.  Viewing the eccentric nuclear disk from the minor axis (middle row), we see that the line-of-sight velocity distributions are generally positively skewed on the left side of the disk near pericenter and negatively skewed on the far right side of the disk near apocenter.

In the right column of Figure~\ref{fig:skew-kurt}, we map the kurtosis in the line-of-sight velocity distributions for an eccentric nuclear disk and a circular disk.  The eccentric nuclear disk generally shows negative values of kurtosis, except in the very center where we see positive kurtosis values.  This is in agreement with observations from \citet{Gultekin2014}, who report mostly negative values of kurtosis for the nuclear stellar disk in NGC 3706.

\begin{figure*}
    \centering
    \includegraphics[trim=3.25cm 3.5cm 3.75cm 4cm, clip=true, width=\textwidth]{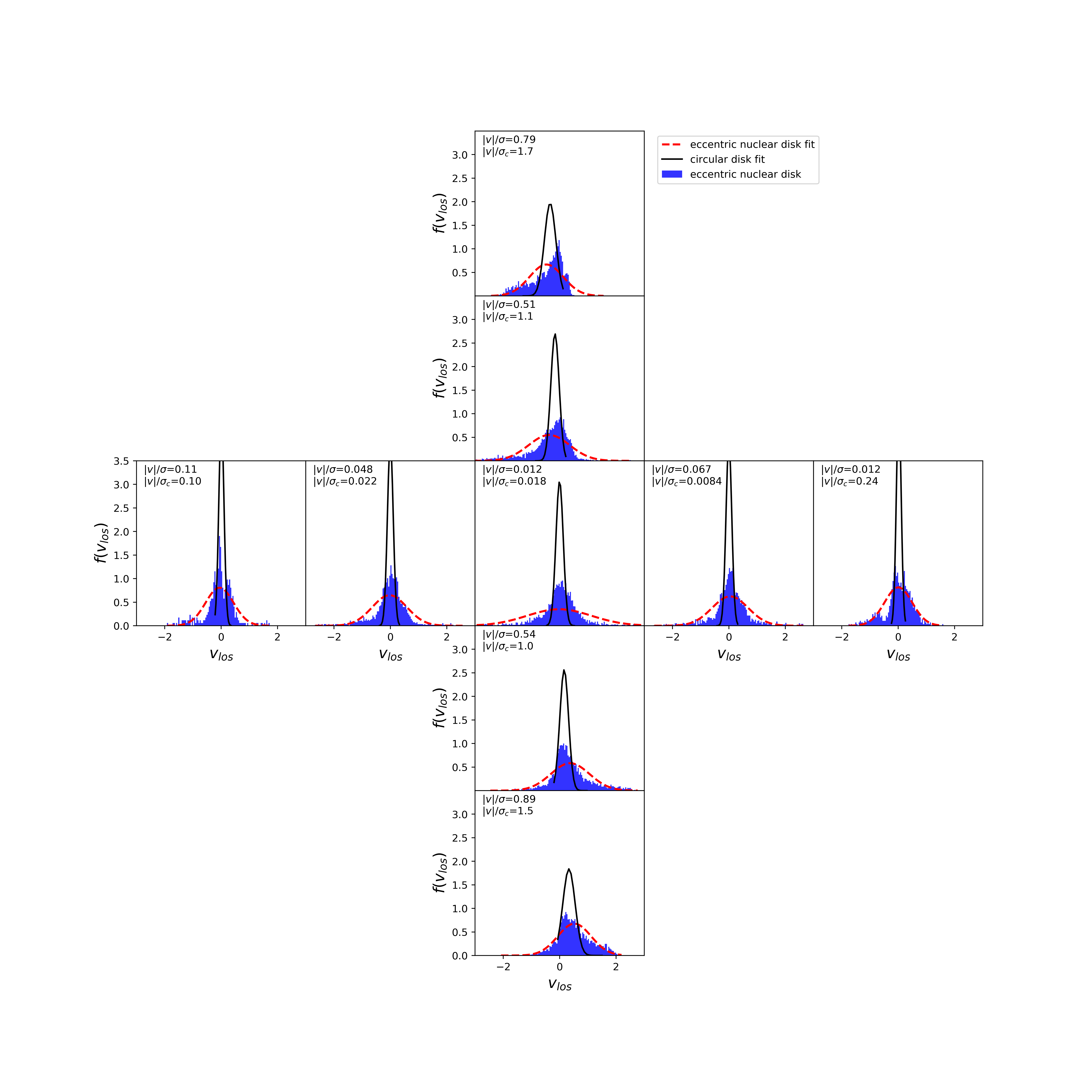}
    \caption{The line-of-sight velocity distributions for single pixels in an eccentric nuclear disk  with N=1000 stars and $0.2$ resolution viewed along its major axis are shown in blue (compare with top panel of Figure~\ref{fig:los-vel}).  Gaussian fits for the eccentric nuclear disk data are shown by the dashed red lines.  
    The solid black lines show Gaussian fits for the line-of-sight velocity distributions in single pixels of a circular disk with N=1000 stars and $0.2$ resolution (compare with bottom panel of Figure~\ref{fig:los-vel}).  
    The pixels are chosen in a cross pattern from the very center of each disk, so that the plot in the very center of the cross depicts the pixel centered on the black hole.  The horizontal row shows five pixels chosen from the center of the disk along x (the minor axis), left to right.  The vertical column shows five pixels chosen from the center of the disk along y, top to bottom.  The velocity distributions for the eccentric nuclear disk are much broader than the narrow distributions of the circular ring.  This means that the quantity $\left|v_{\rm los}\right|/\sigma$ is much lower for the eccentric nuclear disk than would be expected from a circular ring, except for when we look inside the inner edge of the circular disk, where the $v_{\rm los}$ values are extremely small. 
    }
    \label{fig:onepix}
\end{figure*}

In Figure \ref{fig:onepix}, we look at the line-of-sight velocity distribution in several single pixels of the eccentric nuclear disk at {0.2 resolution}.  The pixels are sampled in a cross pattern from the very center (origin) of the eccentric nuclear disk in the top row of Figure~\ref{fig:los-vel}.  We compare the distributions to the distributions of the line-of-sight velocity in the central pixels of the circular disk in the bottom row of Figure~\ref{fig:los-vel}.  In a circular disk, we expect to see a velocity distribution that is peaked at a single value.  Because we take the single pixels from near the origin of the plot, and center of the disk, the single peak should be centered on zero for the circular disk.   
For this reason, the values of $\left|v_{\rm los}\right|/\sigma$ are very small for both the eccentric nuclear disk and the circular disk in the horizontal pixels of Figure~\ref{fig:onepix}.  In the pixels in the vertical portion of the cross however, we see larger values in $\left|v_{\rm los}\right|/\sigma$ for the circular disk and unusually small values for the eccentric nuclear disk.  

\begin{figure*}
    \centering
    \includegraphics[trim=3.25cm 3.5cm 3.75cm 4cm, clip=true, width=\textwidth]{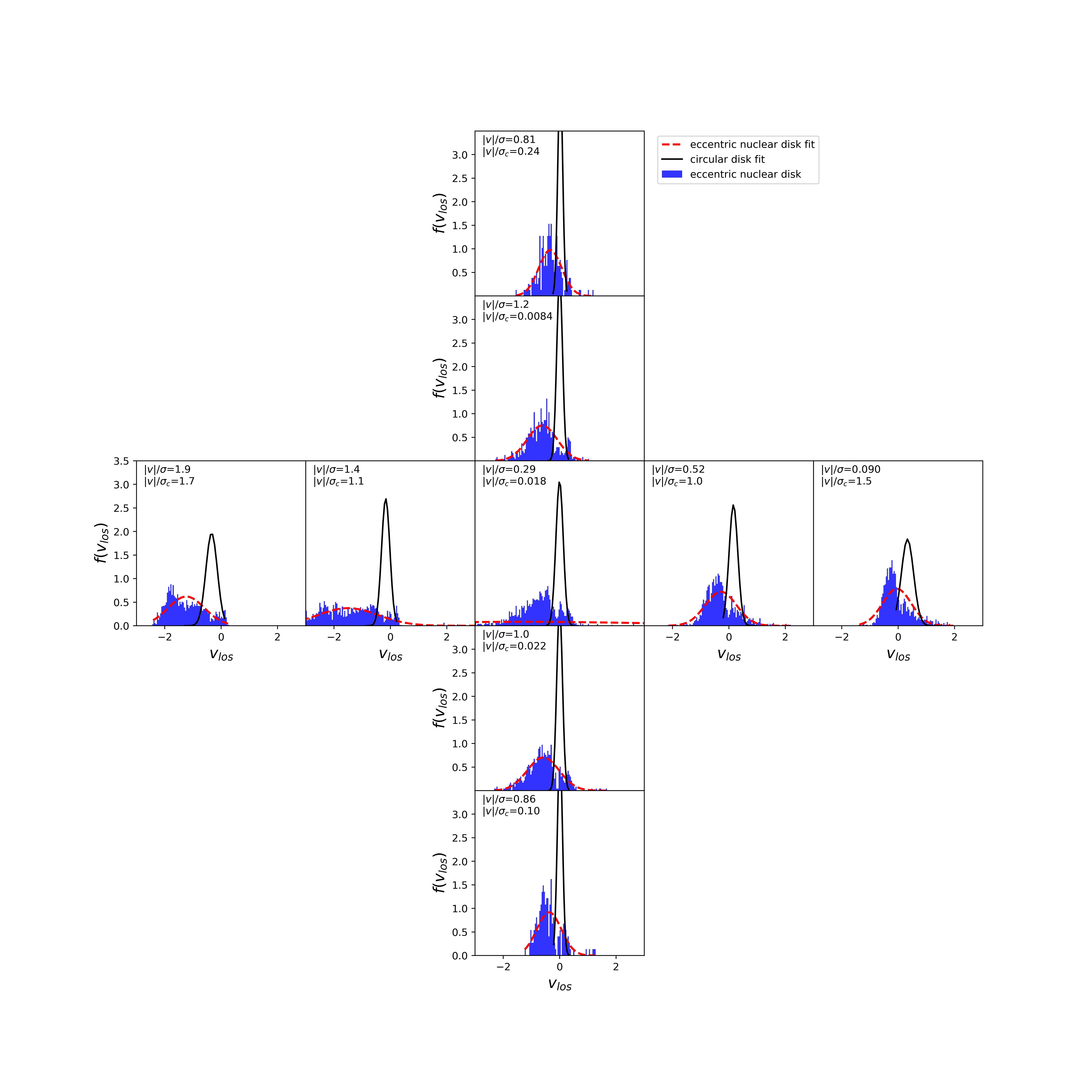}
    \caption{The line-of-sight velocity distributions for single pixels in an eccentric nuclear disk  with N=1000 stars and $0.2$ resolution viewed along its minor axis.  The line-of-sight velocity distributions for the eccentric nuclear disk are shown in blue.  Gaussian fits for the eccentric nuclear disk data are shown by the dashed red lines.  The solid black lines show Gaussian fits for the line-of-sight velocity distribution in single pixels of a circular disk with N=1000 stars, 0.2 resolution, and oriented in the same way as the eccentric nuclear disk in the middle row of Figure~\ref{fig:los-vel}.  The pixels are chosen in a cross pattern from the very center of each disk, so that the plot in the center of the cross depicts the pixel centered on the black hole.  The horizontal row shows five pixels chosen from the center of the disk along the major axis, left to right.  The vertical column shows five pixels chosen from the center of the disk along y, top to bottom.  The velocity distributions for the eccentric nuclear disk are red-shifted and much broader than the narrow distributions for the circular ring.  In the right-most horizontal pixels, the $|v|/\sigma$ values are much smaller in comparison to the circular disk. 
    }
    \label{fig:onepix_minoraxis}
\end{figure*}

Figure~\ref{fig:onepix_minoraxis} looks at the line-of-sight velocity distribution in single pixels of an eccentric nuclear disk, sampling in a cross pattern from the very center (origin) of the eccentric nuclear disk in the middle row of Figure~\ref{fig:los-vel} (viewed along the minor axis).  The distributions are compared to the line-of-sight velocity distributions for a circular disk in the same orientation.  Here, we see that the $\left|v_{\rm los}\right|/\sigma$ values are very small compared to the circular disk in the two right-most pixels in the horizontal portion of the cross.  In the vertical portion of the cross, the $\left|v_{\rm los}\right|/\sigma$ values are so small for the circular disk because the pixels are taken from the inner edge of the disk where the line-of-sight velocities are extremely small.

\section{DISCUSSION}
\label{sec:disc}

In \citet{Madigan2018} and \citet{Wernke2019}, we focused on the internal dynamics of self-gravitating eccentric nuclear disks. Here we look at their photometric and kinematic maps.  We rotate a simulated eccentric nuclear disk to observe the system from different viewing angles.  We classify the resulting observations as double nuclei, centered nuclei, and offset nuclei in comparison to the galaxy sample studied by \citet{Lauer2005}. 
 Out of 77 galaxies, \citet{Lauer2005} found that 12 had centers too dust-obscured to derive surface density photometry. Out of the remaining 65, 5 had an offset nucleus and another 5 had either a double nucleus (1) or a local minimum in their surface brightness (4). That means that $\sim$15\% of galaxies showed evidence for eccentric nuclear disks.  
We also discuss kinematic signatures expected from eccentric nuclear disks.  Our results point to the following conclusions and implications:
\begin{enumerate}
    \item An eccentric nuclear disk with $N=1000$ stars that is viewed from a large number of uniformly sampled angles at a resolution of $0.2$ (where $a = 1$ is the inner edge of the disk) will result in a double nucleus 16\% of the time, an offset nucleus 78\% of the time, and will appear photometrically centered on the black hole 6\% of the time.  As resolution increases, the number of double nuclei observed also increases.  Decreasing the number of simulated stars decreases the fraction of double nuclei observed.
    
    \item \citet{Foote2020} show that more massive bodies in an eccentric nuclear disk segregate to lower orbital inclinations and semi-major axes.  Using their simulation results, we show here that the most massive bodies in an eccentric nuclear disk should preferentially be found in the brighter peak furthest from the supermassive black hole (P1). While this result seems counter-intuitive at first, it is readily explained by the fact that orbits at low semi-major axes and inclination are the most eccentric, and bodies on eccentric orbits linger at apocenter.  The two peaks of an eccentric nuclear disk will have different mass to light ratios.
    
    \item The average line-of-sight velocity values are lower in an eccentric nuclear disk than in a circular disk. 
    The line-of-sight velocity dispersion values are higher  and peak at the position of the supermassive black hole. This does not normally match the peak in photometry which is typically offset from the supermassive black hole in an eccentric nuclear disk.
    
    \item The skew of the line-of-sight velocity distributions of an eccentric nuclear disk generally resemble those of a circular disk (with some differences if the disk is viewed along its minor axis). The kurtosis values are generally negative except in the very center of the disk.
    
    \item The line-of-sight velocity distributions of an eccentric nuclear disk are much broader than the narrow distributions of a circular ring.  In some pixels, 
    the observational quantity $|v_{\rm los}|/\sigma$ is much smaller than expected from a circular ring.  

\end{enumerate}
    
NGC 3706 is an early-type galaxy with a central surface brightness minimum arising from an apparent edge-on stellar ring. Fitting imaging and spectroscopic data to axisymmetric orbit models, \citet{Gultekin2014} uncover a central black hole of mass $M = (6.0 ^{+0.7}_{-0.9}) \times 10^8 M_\odot$. They find however that the stellar ring is inconsistent  with a population of co-rotating stars on circular orbits which would produce a narrow line-of-sight velocity distribution (as in the solid black lines of Figure~\ref{fig:onepix}). Instead the data indicate small line-of-sight values of $|v|/\sigma \sim\!0.1\!\sim\!0.4$. They conclude that the stellar ring contains a retrograde (counter-rotating) component and is not consistent with co-rotating circular orbits. 
We look at this same value in both a simulated circular disk and a simulated eccentric nuclear disk in Figures~\ref{fig:onepix} and ~\ref{fig:onepix_minoraxis} and find similarly low values for the eccentric nuclear disk.
Could this feature be an eccentric nuclear disk? 
For an eccentric nuclear disk to be stable (i.e., apsidally clustered), inter-orbit torques must be sufficiently strong to damp differential apsidal precession. Hence orbits must be well-approximated by closed, Kepler ellipses. This translates to the condition that the disk lies within the radius of influence of a black hole. 
As a consistency check, we determine that the central minimum feature in NGC 3706 lies within the radius of influence of the central black hole.  We use the mass of the supermassive black holes, estimated using the $M-\sigma$ relation \citep{Fer00,Geb00}, to determine the radius of influence \citep{Pee72}. In NGC 3706, the central minimum feature is within the calculated radius of influence.

In this paper we have simulated eccentric nuclear disks in isolation to focus on the effects of disk self-gravity on photometric and kinematic data. 
However, in reality, background stars will significantly contribute to the form of the gravitational potential, thus altering the dynamical evolution of the disk. We briefly discuss how this will affect our results. 
The stability of eccentric nuclear disks relies on the suppression of differential apsidal precession of disk orbits, maintained via inter-orbit gravitational torques \citep{Madigan2018}. Precession also needs to be prograde with respect to the angular momentum of the disk unless the background potential is ``abnormally'' steep \citep{Lyn79, Zderic2021}. If we were to include a typical spherical background stellar distribution in our simulations (due to a cusp or bulge), disk orbits would undergo an enhanced rate of apsidal precession in the {\it retrograde} direction. 
The mass of the disk in our simulations would have to be enhanced to correct for this. In this way \--- including a background stellar potential and enhancing the disk mass \--- would move our simulations in a more astrophysically realistic direction. This is something we are interested in doing in future studies.

In \citet{Wernke2019} we showed that steady-state, eccentric nuclear disks have a non-negligible fraction of retrograde orbiting stars (in our previous simulations this amounted to $\sim$10\%). In our current simulations, where we capture photometric and kinematic maps at 200 orbital periods, we show a lower percentage of retrograde orbits ($\sim$4\%) because the secular dynamical mechanism causing TDEs and retrograde orbits is just beginning to have an effect.  Future work will include analysis of these maps at later times with a larger population of retrograde orbits.  \\


\acknowledgments
We thank Hayden Foote for use of his mass segregation simulation data and the anonymous referee for a thoughtful report. This work was supported by a NASA Astrophysics Theory Program under grant NNX17AK44G. AM gratefully acknowledges support from the David and Lucile Packard Foundation. We utilized the RMACC Summit supercomputer, which is supported by the National Science Foundation (awards ACI-1532235 and ACI-1532236), the University of Colorado Boulder, and Colorado State University. The Summit supercomputer is a joint effort of the University of Colorado Boulder and Colorado State University.

\software{{\tt REBOUND} \citep{Rein2012}}


\bibliography{MasterRefs}

\end{document}